\documentstyle[aps,prl,epsf,floats]{revtex} 
\draft

\begin{document}
\twocolumn[\hsize\textwidth\columnwidth\hsize\csname
@twocolumnfalse\endcsname 


\preprint{UCLA/99/TEP/6}
\title{Highest-energy cosmic rays from Fermi-degenerate relic neutrinos
consistent with Super-Kamiokande results
}

\author{Graciela Gelmini and Alexander Kusenko} 
\address{Department of Physics and Astronomy, UCLA, Los Angeles, CA
90095-1547 } 

\date{February, 1999}

\maketitle
             
\begin{abstract}
Relic neutrinos with mass $0.07^{ +0.02}_{-0.04}$~eV, in the range
consistent with Super-Kamiokande data, can explain the cosmic rays with
energies in excess of the Greisen-Zatsepin-Kuzmin cutoff.  The spectrum of
ultra-high energy cosmic rays produced in this fashion has some distinctive
features that may help identify their origin.  Our mechanism does not
require but is consistent with a neutrino density high enough to be a new
kind of hot dark matter.

\end{abstract}

\pacs{PACS numbers: 98.70.Sa, 95.85.Ry, 14.60.Pq, 95.35.+d \hspace{1.0cm}
UCLA/99/TEP/6 } 

\vskip2.0pc]


\renewcommand{\thefootnote}{\arabic{footnote}}
\setcounter{footnote}{0}


The observation of atmospheric neutrino oscillations at Super-Kamiokande
has provided a strong evidence that at least one of the neutrino species
has mass greater than $m_{_{SK}}=\sqrt{\delta m^2} = 0.07^{
+0.02}_{-0.04}$~eV.  It seems plausible that at least one of the neutrino
masses is actually in this range, as would be the case if the neutrino
masses were hierarchical.  Of course, an alternative possibility, 
that some neutrino masses are nearly degenerate and larger than
$\sqrt{\delta m^2} $, is also consistent with the current data.  In this
paper we will concentrate on the former case.

If one of the lepton asymmetries of the Universe
$L_{i}=(n_{\nu_i}-n_{\bar{\nu}_i})/s$ is of order one~\cite{L0,L1,L2,ls},
the neutrinos with masses $m_{_{SK}}$ can make a significant contribution
to the energy content of the Universe~\cite{pk}.  (Here and below, $n_x$
denotes the number density of the $x$-species, and $s=1.80 g_{s*} n_\gamma$
is the entropy density.) This possibility, frequently discounted in
contemporary cosmology, can arise naturally if the Universe underwent an
Affleck-Dine baryo- and leptogenesis~\cite{ad} at the end of a relatively
low-scale inflation.  Lepton asymmetries can also be generated in neutrino
oscillations~\cite{ftv} after the electroweak phase transition, but we do
not know whether an asymmetry of order one can arise in this fashion.

We will show that ultra-high energy cosmic rays with interesting new
features can be produced in the presence of such background neutrinos.

The cosmic rays with energies beyond the Greisen-Zatsepin-Kuzmin (GZK)
cutoff~\cite{gzk} present a challenging outstanding puzzle in astroparticle
physics and cosmology~\cite{data1,data2}.  The protons with energies above
$5\times 10^{19}$~eV could not reach Earth from a distance beyond 50 -- 100
Mpc~\cite{50Mpc} because they scatter off the cosmic microwave background
photons with a resonant photoproduction of pions: $p\gamma \rightarrow
\Delta^* \rightarrow N\pi$.  The mean free path for this reaction is only
$6$~Mpc.  The photons of comparable energies pair-produce electrons and
positrons on the radio background and, likewise, cannot reach Earth from
beyond 10 to 40 Mpc~\cite{40Mpc}.  This creates a problem because the
closest astrophysical objects that could produce such energetic particles,
active galactic nuclei (AGN), are at least hundreds of megaparsecs away.

Several solutions have been proposed for the origin of these ultra-high
energy cosmic rays (UHECR). For example, they could be produced in the
decay of some ubiquitous hypothetical heavy particles~\cite{particles},
topological defects~\cite{defects}, or light neutral supersymmetric
hadrons~\cite{bf}; or they could also hint at exotic
interactions\cite{domokos}.  A more conservative and economical scenario
involves relic neutrinos.  It has been suggested~\cite{weiler} that distant
AGN's can produce high-energy neutrinos whose annihilation on the relic
neutrinos in the galactic halo at $\sqrt{s} \approx M_{_Z}$ can produce the
protons with energies above the GZK cutoff.  In the absence of lepton
asymmetry, the background neutrino density would not be sufficient to
generate the necessary flux of UHECR if it were not for the clustering of
neutrinos in the galactic halo.  The latter helps.  However, the required
total energy carried by the high-energy neutrinos is uncomfortably close to
the total luminosity of the Universe~\cite{waxman}.

If neutrinos with mass $m_{_{SK}}$ carry a large lepton asymmetry, the
above scenario is aided in several ways.  First, the density $n_\nu$ of the
Fermi-degenerate light neutrinos should be much higher than that considered
in Ref.~\cite{weiler}.  Therefore, the neutrino annihilations in the entire
volume $\sim (50 \ {\rm Mpc})^3$ contribute to the observed UHECR.  Second,
the probability for the neutrinos emitted at distances of the order of the
inverse Hubble constant, $\sim H^{-1}$, to decay within $50$~Mpc of the
observer is maximal when the mean free path $\lambda=1/(\sigma_{\rm ann}
n_\nu)$ is comparable to $H^{-1}$.  We will see that this condition,
$\lambda \sim H^{-1}$, is satisfied automatically for $m_\nu$ measured by
Super-Kamiokande if $\Omega_\nu h^2 \sim 0.01$.  As a result of an
increased background neutrino density and the increase in annihilation
probability in our neighborhood, the total energy of the energetic
neutrinos is much less than the total luminosity of the Universe, in
contrast to Refs.~\cite{weiler,waxman}.  Finally, the predicted spectrum of
the UHECR peaks at $10^{20-21}$~eV and has a cutoff at about $10^{23}$~eV,
hence, creating an observable feature that could help distinguish this
mechanism from some others.

As long as the chemical potential of the degenerate neutrinos $\mu$ is
smaller than the neutrino mass, the neutrinos are non-relativistic.   
Therefore, the neutrinos with mass $m_{_{SK}}$ are non-relativistic at
present if the degeneracy parameter $\xi = \mu/T < 100$.  We will not come
close to this upper bound on $\xi$.   Therefore, the energy density is 
$\rho_i=m_{\nu_i} n_{\nu_i}$ and 
\begin{equation}
\eta_\nu= \frac{n_\nu}{n_\gamma} = 3.6 \left ( \frac{0.07 eV}{m_\nu} \right )
\left ( \frac{\Omega_\nu h^2}{0.01} \right ), 
\label{etaomega}
\end{equation}
where $\Omega_\nu \equiv \rho_i/\rho_c$.  

These neutrinos decouple while they are still relativistic because the
decoupling temperature $T_d$ increases with $\xi$~\cite{ks}, so $T_d>
1$~MeV.  Therefore, the present value of $\xi$ is determined by the
following relation valid for relativistic species:
\begin{equation}
\eta= \frac{1}{12 \zeta(3)} \left ( \frac{T_\nu}{T_\gamma}\right )^3
[\pi^2 \xi + \xi^3] = 0.0252 (9.87 \xi+\xi^3).
\label{etaxi}
\end{equation}
Here we used $\zeta(3)=1.202$ and $(T_\nu/T_\gamma)^3 = 4/11$. This
relation is valid as long as $T_d$ is lower than the muon mass, which
translates into the upper bound $\xi < 12$~\cite{ks}.  From equation
(\ref{etaxi}), $\eta =3.6$, as in equation (\ref{etaomega}), corresponds to
$\xi=4.6$.

The UHECR are dominated by the resonant neutrino annihilations with
$\sqrt{s}\approx M_{_Z}$, which corresponds to the incoming neutrino energy
$E_\nu = M_{_Z}^2/2 m_\nu= 0.57 \times 10^{23}$~eV$(m_{\nu}/0.07 {\rm
eV})$.  The annihilation cross section is $\sigma_{\rm ann} = 4 \pi
G_{_F}/\sqrt{2}$.  The mean free path for a neutrino is, therefore,
\begin{eqnarray}
\lambda  =  \frac{1}{\sigma_{\rm ann} n_\nu} & = & 5.3 
\left ( \frac{412 \, {\rm cm}^{-3}}{n_\nu} \right ) \times 10^{28} {\rm cm}
\nonumber \\ 
& = & \frac{3.9}{ \eta_\nu} 
\left ( \frac{0.65}{h} \right) H^{-1} . 
\label{lambda} 
\end{eqnarray}

The atmospheric neutrino oscillations observed at Super-Kamiokande imply
that a muon neutrino has a large, order one, mixing with either $\nu_\tau$
or a sterile neutrino.  The mass eigenstate that makes up the
Fermi-degenerate relic background with mass $m_{_{SK}}$ must, therefore,
have a muon neutrino component that is not small.  The astrophysical
sources are expected to produce high-energy muon neutrinos from the decays
of pions.  Since both the background neutrinos and the high-energy
neutrinos from astrophysical sources must have a large muon component,
there is no significant suppression of the annihilation cross section due
to mixing angles.

The probability of the neutrino annihilation is maximized when $\eta_\nu
\simeq 4$, at which point a fraction $r=(50 \ {\rm Mpc}/2.7 \lambda)
\approx 5.5 \times 10^{-3}$ of all neutrinos with energies near the $Z$
resonance annihilate within 50~Mpc of the observer.  According to the data
pertaining to the momentum distribution of $Z$ decay products~\cite{pdg},
these neutrino annihilations yield hadrons with an average energy
\begin{equation}
E_p \sim  0.025
E_\nu \approx 1.4 \left ( \frac{0.07 \ {\rm eV}}{m_\nu} \right ) 
\times 10^{21} {\rm eV}
\label{Ep}
\end{equation}
and photons (from the decays of $\pi^0$) with a lower average energy, 
\begin{equation}
E_\gamma \sim 0.0035 E_\nu \approx 2.0 \left ( \frac{0.07 \ {\rm
eV}}{m_\nu} \right ) \times 10^{20} {\rm eV}.
\label{Eg}
\end{equation}
$Z$ decays produce on average 2 nucleons and about 9.5 $\pi^0$'s, which
decay into 19 photons\cite{pdg}.  Even though the photons that reach Earth
originate in a smaller volume (10--40 Mpc)$^3$ than protons (50-100
Mpc)$^3$, both components should contribute to the UHECR because the
multiplicity of photons is about 10 times greater than that of protons.  It
may be possible, given enough statistics of UHECR, to resolve the two peaks
in the distribution of cosmic rays: one at lower mean energy, due to
photons, and another, at higher energies, due to protons.

The total energy per unit volume in neutrinos with energies above
$10^{19}$~eV in our case is lower than that in Refs.~\cite{weiler,waxman}
by a factor $21 (\eta_\nu/4)( m_\nu/0.07 {\rm eV})$.  The difference with
$E(>10^{19}{\rm eV})$ in equation (5) of Ref.~\cite{waxman} is the increased
value of the background neutrino density, $m_\nu = m_{_{SK}}$, and
$N_{_{CR}} \sim 30$.  This means the total power generated in high-energy
neutrinos ${\cal E}_\nu \sim 0.5 (4/\eta_\nu) (0.07 {\rm eV}/ m_\nu)
10^{48}{\rm erg \: Mpc}^{-1} {\rm yr}^{-1}$ is well below the luminosity of
the Universe.

Bounds on the neutrino degeneracy come from
nucleosynthesis~\cite{ks,nucleo}, as well as structure formation in the
Universe~\cite{fk}.  A combination of both yields $-0.06
\stackrel{<}{_{\scriptstyle \sim}} \xi_{\nu_e} \stackrel{<}{_{\scriptstyle
\sim}} 1.1$, $|\xi_{\nu_{\mu,\tau}}| \stackrel{<}{_{\scriptstyle \sim}}
6.9$~\cite{ks}.  In addition, in models with large neutrino degeneracy, the
baryon density of the Universe can be higher than in conventional
nucleosynthesis, so a larger fraction of the dark matter can be baryonic.
These bounds are based on the requirement that neutrinos not interfere with
galaxy formation.  However, it has been shown recently~\cite{subir} that a
relativistic degenerate neutrino may actually help the formation of
structure in the Universe.  This neutrino may be the addition that
`standard' cold dark matter (CDM) models need to account well for all
present data on structure in the Universe.  In fact, a CDM model with a
relativistic relic neutrino background with $\xi \sim 3.4$~\cite{subir}
provides a good fit to all the data on large-scale structure and anisotropy
of the cosmic microwave background radiation.  These analyses do not fully
apply to our case because the neutrinos with mass $m_{_{SK}}$ are
non-relativistic at present.  Our scenario is consistent with large values
of $\Omega_\nu h^2$ that can make the relic degenerate neutrinos an
important hot dark matter component.  Its effects on structure formation
and the anisotropy of CMBR need to be studied.

Finally, we would like to address the issue of how the large lepton
asymmetry could be generated in the early Universe, while the baryon
asymmetry remains small.  The coherent Affleck-Dine condensate could have
evolved differently along the flat directions carrying the baryon ($B$)
and the lepton ($L$) numbers.  The corresponding supersymmetry-breaking
terms and higher-dimension operators~\cite{ad_flat} depend on the type of a
flat direction and need not be the same for the directions with $B\neq 0$
and $L\neq 0$.  If the electroweak symmetry was never restored after
inflation, either because the reheat temperature was low, or because the
lepton asymmetry was high~\cite{L1,ls}, the baryon and lepton asymmetries
of the Universe may differ by many orders of magnitude.  A cold dark matter
component, called for by the need to form structure, could also arise
naturally in the low-scale Affleck-Dine scenario~\cite{dm_from_ad}.  In
this paper we do not assume any particular cosmological scenario for
generating the relic neutrinos.  However, it is reassuring that an
economical self-consistent cosmological model outlined above can
simultaneously produce cold and hot dark matter, the latter being the
light, Fermi-degenerate neutrinos that carry a lepton asymmetry of order
one.

To summarize, we have shown that a cosmic background of Fermi-degenerate
neutrinos with masses inferred from the Super-Kamiokande data can explain
the ultra-high energy cosmic rays above the GZK cutoff.  Our mechanism does
not require but is consistent with a neutrino density high enough to be a
new kind of hot dark matter.  The mechanism predicts ultra-high energy
cosmic rays from both protons, peaked at energies $10^{22}$~eV, and
photons, peaked at $10^{20}$~eV.  Another prediction is a new cutoff at 
$M_{_Z}^2/2 m_\nu= 0.57 \times 10^{23}$~eV$(m_{\nu}/0.07 {\rm eV})$. 

This work was supported in part by the US Department of Energy grant
DE-FG03-91ER40662, Task C.



\begin{references}

\bibitem{L0} A.~D.~Dolgov and D.~P.~Kirilova, J. Moscow Phys. Soc. {\bf 1},
217 (1991); A.~D.~Dolgov, Phys. Rep. {\bf 222}, 309 (1992).

\bibitem{L1} A.~Casas, W.~Y.~Cheng, and G.~Gelmini, Nucl. Phys. {\bf B538},
297 (1999).

\bibitem{L2} J.~A.~Harvey and E.~W.~Kolb, Phys. Rev. {\bf D24}, 2090
(1981); J.~Liu and G.~Segr\`e, Phys. Rev. {\bf D48}, 4609 (1993); B.~Bajc,
A.~Riotto, and G.~Senjanovi\'c, Phys. Rev. Lett. {\bf 81}, 1355 (1998). 


\bibitem{ls} A.~Linde, Phys. Rev. {\bf D14}, 3345 (1976); P.~Langacker,
 G.~Segr\`e, and S.~Soni, Phys. Rev. {\bf D26}, 3425 (1982); J.~Liu and
 G.~Segr\`e, Phys. Lett. {\bf B338}, 259 (1994).

\bibitem{pk} P.~Pal and K.~Kar, hep-ph/9809410.  

\bibitem{ad} I.~Affleck and M.~Dine, Nucl. Phys. {\bf B249}, 361 (1985). 

\bibitem{ftv} R.~Foot, M.~J.~Thompson, and R.~R.~Volkas, Phys. Rev. {\bf
D53}, 5349 (1996); X.~Shi, Phys. Rev. {\bf D54}, 2753 (1996).

\bibitem{gzk} K.~Greisen, {\sl Phys. Rev. Lett.} {\bf 16}, 748 (1966);
   G.~T.~Zatsepin and V.~A.~Kuzmin, {\sl Pisma Zh. Eksp. Teor. Fiz.} {\bf
   4}, 114 (1966).

\bibitem{data1} D.~J.~Bird {\it et al.}, Phys. Rev. Lett. {\bf 71}, 3401
(1993); Astrophys. J. {\bf 424}, 491 (1994).

\bibitem{data2} S.~Yoshida {\it et al.}, Astropart. Phys. 3, 151 (1995). 

\bibitem{50Mpc} S.~Yoshida and M.~Teshima, Prog. Theor. Phys.  {\bf
        89}, 833 (1993); F.~A.~Aharonian and J.~W.~Cronin, {
        Phys. Rev. } {\bf D50}, 1892 (1994); J.~W.~Elbert and P.~Sommers,
        {Astrophys. J.} {\bf 441}, 151 (1995);

\bibitem{40Mpc} F.~Halzen, R.~A.~Vazquez, T.~Stanev, and V.~P.~Vankov, 
  Astropart. Phys., {\bf 3}, 151 (1995).

\bibitem{particles} V.~Berezinsky, M.~Kachelriess, and A.~Vilenkin,
Phys. Rev. Lett. {\bf 79}, 4302 (1997); V.~A.~Kuzmin and V.~A.~Rubakov,
Phys. Atom. Nucl. {\bf 61}, 1028 (1998) [Yad. Fiz. {\bf 61}, 1122 (1998)];
M.~Birkel and S.~Sarkar, Astropart. Phys. {\bf 9}, 297 (1998); V.~Kuzmin
and I.~Tkachev, JETP Lett. {\bf 68}, 271 (1998); K.~Benakli, J.~Ellis, and
D.~V.~Nanopoulos, Phys. Rev. {\bf D59}, 047301 (1999).

\bibitem{defects} C.~T.~Hill, D.~N.~Schramm, and T.~P.~Walker,
Phys. Rev. {\bf D36}, 1007 (1987); G.~Sigl, D.~N.~Schramm and
P.~Bhattacharjee, Astropart. Phys. {\bf 2}, 401 (1994); V.~Berezinsky,
X.~Martin, and A.~Vilenkin, Phys.Rev. {\bf D56}, 2024 (1997).

\bibitem{bf} D.~J.~Chung, G.~R.~Farrar, and E.~W.~Kolb, Phys. Rev. {\bf
D57}, 4606 (1998); G.~R.~Farrar and P.~Biermann, Phys. Rev. Lett. {\bf 81},
3579 (1998); I.~F.~Albuquerque, G.~R.~Farrar, and E.~W.~Kolb,
Phys. Rev. {\bf D59}, 015021 (1999).

\bibitem{domokos} G.~Domokos and S.~Kovesi-Domokos, Phys. Rev. Lett. {\bf
82}, 1366 (1999).

\bibitem{weiler} D.~Fargion, B.~Mele, and A.~Salis, astro-ph/970029;
T.~Weiler, hep-ph/9710431. 
 
\bibitem{waxman} E.~Waxman, astro-ph/9804023.  

\bibitem{ks} H.~Kang and G.~Steigman, Nucl. Phys. {\bf B372}, 494 (1992). 

\bibitem{pdg} Particle Data Group, Eur. Phys. J. {\bf C3}, 201-202 (1998). 

\bibitem{nucleo} R.~V.~Wagoner, W.~A.~Fowler and F.~Hoyle, Ap. J. {\bf
148}, 3 (1967); A. Yahil and G.~Beaudet, Ap. J. {\bf 206}, 26 (1976);
Y.~David and H.~Reeves, {\it Physical Cosmology}, ed. by R.~Balian,
J.~Audouze, and D.~N.~Schramm, North-Holland, Amsterdam, 1980; J.~N.~Fry
and C.~J.~Hogan, Phys. Rev. Lett. {\bf 49}, 1783 (1982); R.~J.~Scherrer,
Mon. Not. Roy. Astron. Soc. {\bf 205}, 683 (1983); N.~Terazawa and K.~Sato,
Ap. J. {\bf 294}, 9 (1985); K. Olive, D.~N.~Schramm, D.~Thomas and
T.~Walker, Phys.~Lett. {\bf B265}, 239 (1991); G.~Starkman Phys. Rev. {\bf
D45}, 476 (1992).

\bibitem{fk} K.~Freese, E.~W.~Kolb, and M.~S.~Turner, Phys. Rev. {\bf D27},
1689 (1983).

\bibitem{subir} J.~Adams and S.~Sarkar, talk presented at the {\it ICTP
Workshop on physics of relic neutrinos}, Trieste, Italy, 1998.

\bibitem{ad_flat} M.~Dine, L.~Randall, and S.~Thomas, Phys. Rev. Lett. {\bf
75}, 398 (1995); Nucl. Phys. {\bf B458}, 291 (1996); J.~A.~Casas and
G.~B.~Gelmini, Phys. Lett. {\bf B410}, 3 (1997); B.~A. Campbell,
M.~K.~Gaillard, H.~Murayama, and K.~A.~Olive, Nucl. Phys. {\bf B538}, 351
(1999); M.~Axenides, E.~G.~Floratos, G.~K.~Leontaris, and N.~D.~Tracas,
hep-ph/9811371.

\bibitem{dm_from_ad} A.~Kusenko and M.~Shaposhnikov, Phys. Lett. {\bf
  B418}, 46 (1998); K.~Enqvist and J.~McDonald, Phys. Lett. {\bf B 425},
  309 (1998).

\end{references}
\end{document}